\documentclass[bib]{ba}

\usepackage{graphicx,hyperref}
\usepackage{amssymb,amsfonts,amsmath}
\usepackage{natbib}
\usepackage{appendix}
\usepackage{array,booktabs}
\usepackage[standard]{ntheorem}
\usepackage[algo2e]{algorithm2e}

\theoremstyle{break}
\theorembodyfont{\normalfont}
\newtheorem{algorithm}[algocf]{Algorithm}

\usepackage{bayes}

\begin{document}


\inserttype[ba0001]{article}
\renewcommand{\thefootnote}{\fnsymbol{footnote}}
\author{Mohammad Arshad Rahman}{
 \fnms{Mohammad Arshad}
 \snm{Rahman}
 \footnotemark[1]\ead{email1@example.com}
}

\title[BQR for Ordinal Models]{Bayesian Quantile Regression for Ordinal Models}

\maketitle

\footnotetext[1]{Assistant Professor of Economics, Department of Humanities
and Social Sciences, Indian Institute of Technology Kanpur, India.
 \href{mailto:email1@example.com}{mailto:marshad@iitk.ac.in}
}
\renewcommand{\thefootnote}{\arabic{footnote}}

\begin{abstract}
The paper introduces a Bayesian estimation method for quantile regression
in univariate ordinal models. Two algorithms are presented that utilize the
latent variable inferential framework of \citet{Albert-Chib-1993} and the
normal-exponential mixture representation of the asymmetric Laplace
distribution. Estimation utilizes Markov chain Monte Carlo simulation --
either Gibbs sampling together with the Metropolis-Hastings algorithm or
only Gibbs sampling. The algorithms are employed in two simulation studies
and implemented in the analysis of problems in economics (educational
attainment) and political economy (public opinion on extending ``Bush Tax''
cuts). Investigations into model comparison exemplify the practical utility
of quantile ordinal models.

\end{abstract}

\section{Introduction}

Quantile regression \citep{Koenker-Basset-1978} models the relationship
between  the covariates and the conditional quantiles of the dependent
variable. The methodology supplements least squares regression and provides a
more comprehensive picture of the underlying relationships of interest that
can be especially useful when relationships in the lower or upper tails are
of significant interest. Estimation of quantile regression models require
implementation of specialized algorithms and reliable estimation techniques
have been developed in both the classical and Bayesian literatures, primarily
for cases when the dependent variable is continuous. Classical techniques
include the simplex algorithm
\citep{Dantzig-1963,Dantzig-Thapa-1997,Dantzig-Thapa-2003,Barrodale-Roberts-1973,Koenker-dOrey-1987},
and the interior point algorithm
\citep{Karmarkar-1984,Mehrotra-1992,Portnoy-Koenker-1997}, whereas Bayesian
methods relying on Markov chain Monte Carlo (MCMC) sampling have been
proposed in \citet{Yu-Moyeed-2001}, \citet{Tsionas-2003},
\citet{Reed-Yu-2009}, and \citet{Kozumi-Kobayashi-2011}.

The advantage of quantile regression, as a more informative description of
the relationships of interest, also applies to models where the dependent
variable is discrete and ordered, i.e. `ordinal models'. Ordinal models are
very common and arise in a wide class of applications across disciplines
including business, economics, political economy and the social sciences.
However, the literature does not offer many alternatives when it comes to
quantile estimation of ordinal models. The difficulties stem from the
nonlinearity of the link function, the discontinuity of the loss function and
the location and scale restrictions required for parameter identification. In
the classical literature, estimation of quantile regression in ordinal models
has been addressed only in the last few years. \citet{Zhou-2010} adopted the
latent variable approach and estimated quantile regression with ordinal data
using simulated annealing \citep{Kirkpatrick-etal-1983,Goffe-etal-1994}.
\citet{Hong-He-2010} developed the transformed ordinal regression quantile
estimator (TORQUE) for single-index semiparametric ordinal models and showed
that TORQUE could be used to produce conditional quantile estimates and
construct prediction intervals. Although useful, the approach has the
practical limitation of requiring the assumption of zero correlation between
the errors and the single-index. The drawback was addressed by
\citet{Hong-Zhou-2013}, who introduced a multi-index model to explicitly
account for any remaining correlation between the covariates and the
residuals from the single-index model. In contrast, Bayesian techniques for
estimating quantile ordinal models have not yet been proposed.

The paper fills the above mentioned gap and introduces MCMC algorithms for
estimating quantile regression in ordinal models. The proposed method
utilizes the latent variable inferential framework of
\citet{Albert-Chib-1993} together with the normal-exponential mixture
representation of the asymmetric Laplace (AL) distribution
\citep[see][]{Kotz-etal-2001,Yu-Zhang-2005}. The normal-mixture
representation is employed because it offers access to the convenient
properties of the normal distribution and simplifies the sampling process.
Location and scale restrictions are enforced by anchoring a cut-point and
fixing either the error variance or a second cut-point, respectively. The
paper shows that judicious use of the scale restriction can play an important
role in simplifying the sampling procedure. The algorithms are illustrated in
two simulation studies and employed in two applications involving educational
attainment and public opinion on the extension of ``Bush Tax'' cuts by
President Obama. Both applications provide interesting results and raise
suggestions for future work. In addition, model comparison using deviance
information criterion (DIC) shows that quantile ordinal models can provide a
better model fit as compared to the commonly used ordinal probit model. Note
that the objective is only to compare and contrast the proposed models with
an ordinal probit model, but they should not be used as substitutes since
they are different models and focus on different quantities, i.e. quantiles
as opposed to mean.

The remainder of the paper is organized as follows. Section \ref{sec:QR}
introduces the quantile regression problem and its formulation in the
Bayesian context. Section \ref{sec:QRinORmodels} presents the quantile
ordinal model and discusses estimation procedures together with Monte Carlo
simulation studies. Section \ref{sec:applications} presents the applications
and Section \ref{sec:conclusion} concludes.

\section{Quantile Regression}\label{sec:QR}

The $p$-{th} quantile of a random variable $Y$  is the value $y_{0}$ such
that the probability that $Y$ will be less than $y_{0}$ equals $p \in (0,1)$.
Formally, if $F(\cdot)$ denotes the cumulative distribution function of $Y$,
the $p$-{th} quantile is defined as
\begin{equation*}
F^{-1}(p) =
\mathrm{inf} \{y_{0}: F(y_{0})\geq p\}.
\end{equation*}
The idea of quantiles is extended to regression analysis via quantile
regression, where the aim is to estimate \emph{conditional quantile
functions} with $F( \cdot)$ being the conditional distribution function of
the dependent variable given the covariates. An interesting feature of
quantile regression is that the quantile objective function is a sum of
asymmetrically weighted absolute residuals, minimization of which yields
regression quantiles.

\begin{figure*}[!t]
  \centerline{
    \mbox{\includegraphics[width=3.00in, height=2.00in]{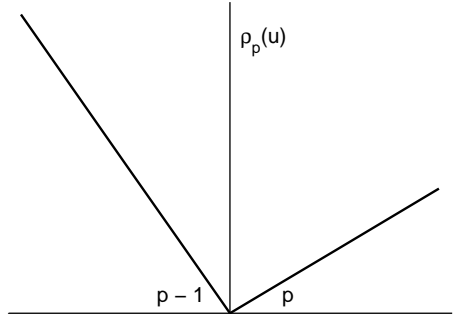}}
    }
\caption{Quantile regression check function.} \label{fig:CheckFunction}
\end{figure*}

In order to formally explain the quantile regression problem, consider a
linear model,
\begin{equation}
y = X \beta_{p} + \epsilon,
\label{eq:model1}
\end{equation}
where $y$ is an $n\times 1$ vector of responses, $X$ is an $n \times k$
covariate matrix, $\beta_{p}$ is a $k \times 1$ vector of unknown parameters
that depend on quantile $p$ and $\epsilon$ is an $n \times 1$ vector of
unknown errors. In the classical literature, the error distribution is not
specified and quantile regression estimation proceeds by minimizing, with
respect to $\beta_{p}$, the following objective function,
\begin{equation}
\min_{\beta_{p} \in \mathbf{R}^{k}} \bigg[\;  \sum_{i:y_{i}
      < x_{i}'\beta_{p}} (1 - p)\; |y_{i} - x_{i}'\beta_{p}| \;\;
      + \sum_{i:y_{i} \geq x_{i}'\beta_{p}}  p \; |y_{i} - x_{i}'\beta_{p}|
      \;\; \bigg],
      \label{eq:objfunc}
\end{equation}
where the solution vector $\hat{\beta}_{p}$ gives the $p$-th regression
quantile and the estimated conditional quantile function is obtained as $
\hat{y} = X \hat{\beta}_{p}$. Note that the objective function
\eqref{eq:objfunc} is such that all observations above the estimated
hyperplane $X\hat{\beta}_{p}$ are weighted by $p$, and all observations below
the estimated hyperplane are weighted by $(1 - p)$. Therefore, the objective
function \eqref{eq:objfunc} can be written as a sum of check functions or
piecewise linear functions as follows, \begin{equation*} \min_{\beta_{p} \in
\mathbf{R}^{k}}  \sum_{i=1}^{n} \rho_{p}(y_{i} - x'_{i}\beta_{p}),
\end{equation*} where $\rho_{p}(u) = u \cdot (p - I(u<0))$ and $I(\cdot)$ is
an indicator function, which equals 1 if the condition inside the parenthesis
is true and 0 otherwise. It is obvious from Figure~\ref{fig:CheckFunction}
that the check function is not differentiable at the origin and consequently,
classical methods rely upon computational techniques such as the simplex
algorithm, the interior point algorithm or the smoothing algorithm
\citep{Madsen-Nielsen-1993, Chen-2007}. Simulation methods, such as
metaheuristic algorithms, can also be used to estimate quantile regression
models \citep{Rahman-2013}.

The Bayesian method of estimating quantile regression uses the fact that
maximization of the likelihood, where the error follows an AL distribution,
is equivalent to minimization of the quantile objective function
\citep{Yu-Moyeed-2001}. The error $\epsilon_{i}$ follows a skewed AL
distribution, denoted $\epsilon_{i} \sim AL(0,1,p)$, if the probability
density function (pdf) is given by:
\begin{equation}
f_{p}(\epsilon_{i}) = p(1-p)\left\{ \begin{array}{ll}
         \exp(-\epsilon_{i} \,(p-1)) & \mathrm{if \; \epsilon_{i} <0}   \label{eq:laplace}\\
         \exp(-\epsilon_{i} \,p)     & \mathrm{if \; \epsilon_{i} \geq 0},
\end {array} \right.
\end{equation}
where the location, scale and skewness parameters equal 0, 1, and $p$,
respectively \citep{Kotz-etal-2001,Yu-Zhang-2005}. The mean and variance of
$\epsilon_{i}$ with pdf \eqref{eq:laplace} are as follows,
\begin{equation*}
E(\epsilon_{i}) = \frac{1 - 2p}{p(1-p)} \quad \textrm{and} \quad
V(\epsilon_{i}) = \frac{1-2p+2p^{2}} {p^{2}(1-p)^{2}}.
\end{equation*}
Both mean and variance, as shown above, depend on the skewness parameter $p$,
but are fixed for a given value of $p$. Interestingly, $p$ also defines the
quantile of an AL distribution and the $p$-th quantile is always zero. This
feature becomes useful in quantile regression, since estimation of a model at
different quantiles simply requires a change in the value of $p$.

Given the likelihood based on the AL distribution, the posterior distribution
is proportional to the product of the likelihood and the prior distribution
of the parameters. Unfortunately, the joint posterior distribution does not
have a known tractable form and typically requires MCMC methods for posterior
inferences. In this context, \citet{Kozumi-Kobayashi-2011} show that Gibbs
sampling can be employed provided the AL distribution is represented as a
mixture of normal-exponential distributions, i.e.
\begin{equation}
\epsilon_{i} = \theta w_{i} + \tau \sqrt{w_{i}} \, u_{i},  \hspace{0.75in}
\forall \; i=1,\cdots,n,
\label{eq:normal-exp}
\end{equation}
where $w_{i}$ and $u_{i}$ are mutually independent, $u_{i} \sim N(0,1)$,
$w_{i} \sim\mathcal{E}(1)$, and $\mathcal{E}$ represents an exponential
distribution. The constants ($\theta, \tau$) in equation
\eqref{eq:normal-exp} are defined as follows,
\begin{equation*}
\theta = \frac{1-2p}{p(1-p)} \qquad \mathrm{and} \qquad \tau = \sqrt{\frac{2}{p(1-p)}}.
\end{equation*}
The normal-exponential mixture representation of the AL distribution offers
access to properties of the normal distribution, which are exploited in the
current paper to derive the sampler for quantile regression in ordinal
models.

\section{Quantile Regression in Ordinal Models}\label{sec:QRinORmodels}

Ordinal models arise when the dependent (response) variable is discrete and
outcomes are inherently ordered or ranked with the characteristic that scores
assigned to outcomes have an ordinal meaning, but no cardinal interpretation.
For example, in a survey regarding the performance of the economy, responses
may be recorded as follows: 1 for `bad', 2 for `average' and 3 for `good'.
The responses in such a case have ordinal meaning but no cardinal
interpretation, so one cannot say a score of 2 is twice as good as a score of
1. The ordinal ranking of the responses differentiates these data from
unordered choice outcomes.

A quantile regression ordinal model can be represented using a continuous
latent random variable $z_{i}$ as,
\begin{equation}
z_{i} = x'_{i} \beta_{p}  + \epsilon_{i}, \hspace{0.75in} \forall \; i=1, \cdots, n,
\label{eq:model2}
\end{equation}
where $x_{i}$ is a $k \times 1$ vector of covariates, $\beta_{p}$ is a $k
\times 1$ vector of unknown parameters at the $p$-th quantile, $\epsilon_{i}$
follows an AL distribution with pdf \eqref{eq:laplace} and $n$ denotes the
number of observations. However, the variable $z_{i}$ is unobserved and
relates to the observed discrete response $y_{i}$, which has \textit{J}
categories or outcomes, via the cut-point vector $\gamma_{p}$ as follows:
\begin{equation}
\gamma_{p,j-1} < z_{i} \leq \gamma_{p,j} \; \Rightarrow \; \emph{$y_{i}$ = j}, \hspace{0.75in}
                                             \forall \; i=1,\cdots, n; \; j=1,\cdots, J,
                                             \label{eq:cutpoints}
\end{equation}
where $\gamma_{p,0}=-\infty$ and $\gamma_{p,J}=\infty$. In addition,
$\gamma_{p,1}$ is typically set to 0, which anchors the location of the
distribution required for parameter identification (see
\citealp{Jeliazkov-etal-2008}). Given the data vector $y$ = $(y_{1}, \cdots,
y_{n})'$, the likelihood for the model expressed as a function of unknown
parameters $(\beta_{p}, \gamma_{p})$ can be written as,
\begin{equation}
\begin{split}
f(\beta_{p}, \gamma_{p}; y)
      &  =  \prod_{i=1}^{n} \prod_{j=1}^{J} P(y_{i} = j | \beta_{p},
            \gamma_{p})^{ I(y_{i} = j)}  \\
      &  =  \prod_{i=1}^{n}  \prod_{j=1}^{J}
            \bigg[ F_{AL}(\gamma_{p,j} - x'_{i}\beta_{p}) -
            F_{AL}(\gamma_{p,j-1} - x'_{i}\beta_{p})   \bigg]^{ I(y_{i} = j)}
            \label{eq:likelihood}
\end{split}
\end{equation}
where, $F_{AL}(\cdot)$ denotes the cumulative distribution function
(\emph{cdf}) of an AL distribution and $I(y_{i}=j)$ is an indicator function,
which equals 1 if $y_{i}=j$ and 0 otherwise.

The Bayesian approach to estimating quantile ordinal models utilizes the
latent variable representation \eqref{eq:model2} together with the
normal-exponential representation \eqref{eq:normal-exp} of the AL
distribution. The $p$-th quantile ordinal model can therefore be expressed
as,
\begin{equation}
z_{i} = x'_{i} \beta_{p}  + \theta w_{i} + \tau \sqrt{w_{i}} \,u_{i},
        \hspace{0.5in} \forall \; i=1, \cdots, n.
\label{eq:model3}
\end{equation}
It is clear from formulation \eqref{eq:model3} that the latent variable
$z_{i}|\beta_{p},w_{i} \sim N( x'_{i}\beta_{p} + \theta w_{i}, \tau^{2}
w_{i})$, allowing usage of the convenient properties of normal distribution
in the estimation procedure.

Before moving forward, it is beneficial to subdivide ordinal models, as
$\mathrm{OR_{I}}$ and $\mathrm{OR_{II}}$, based on the number of outcomes and
the type of scale restriction employed. The subdivision is employed to
present two algorithms --- a general algorithm for estimation of ordinal
models that utilizes Gibbs sampling and the Metropolis-Hastings (MH)
algorithm, and a simpler algorithm for estimation of ordinal models with
three outcomes that solely relies on Gibbs sampling. They form the subject of
discussion in the next two subsections.

On a side note, although the derivation of posterior distributions for
$\mathrm{OR_{I}}$ and $\mathrm{OR_{II}}$ models utilize a normal prior on
$\beta_{p}$, it is not the default choice. One may also employ the
normal-exponential mixture representation of the Laplace or double
exponential distribution as the prior distribution
\citep{Andrews-Mallows-1974, Park-Casella-2008, Kozumi-Kobayashi-2011}. The
full conditional posteriors with a Laplace prior is a straightforward
modification of the derivations with a normal prior, and hence has not been
presented to keep the paper within reasonable length.

\subsection{$\mathrm{OR_{I}}$ Model} \label{sec:ORI}

The term ``$\mathrm{OR_{I}}$ model,'' as used in the paper, refers to an
ordinal model (equations (\ref{eq:model3}) and (\ref{eq:cutpoints})) in which
the number of outcomes is greater than three ($J > 3$), location restriction
is achieved \emph{via} $\gamma_{p,1} = 0$ and scale restriction is enforced
\emph{via} fixed variance (since $V(\epsilon_{i})$ is constant for a given
$p$). The location restriction removes the possibility of shifting the
distribution without changing the probability of observing $y_{i}$ and the
scale restriction fixes the scale of the latent data that is implied by the
\emph{cdf} of the AL distribution, i.e., $F_{AL}$. Note that one may
incorporate $J=3$ outcomes within the definition of $\mathrm{OR_{I}}$ model,
but estimation would involve the MH algorithm, which can be avoided as
presented in Section \ref{sec:ORII}.

\subsubsection{Estimation} \label{sec:ORIest}

Estimation of the $\mathrm{OR_{I}}$ model utilizes the approach of
\citet{Kozumi-Kobayashi-2011} with the addition of the following two
components: first, location and scale restrictions, and second, threshold or
cut-point vector $\gamma_{p}$. The location and scale restrictions are easy
to impose, but sampling of the cut-points requires additional consideration.
In particular, two issues arise with respect to the sampling of $\gamma_{p}$:
the ordering constraints and the absence of a known conditional distribution
of the transformed cut-points.

The ordering constraints within the cut-point vector $\gamma_{p}$ cause
complication since it is difficult to satisfy the ordering during sampling.
Ordering can be removed by using any monotone transformation from a compact
set to the real line. The paper employs the logarithmic transformation,
\begin{equation}
\delta_{p,j} = \ln ( \gamma_{p,j} - \gamma_{p,j-1} ), \qquad 2 \le j \le J-1.
\label{eq:logtransformation}
\end{equation}
Other transformations, such as log-ratios of category bin-widths or
trigonometric functions like arctan and arcsin, are also possible. The
original cut-points can then be obtained by a one-to-one mapping between
$\delta_{p} = (\delta_{p,2}, \cdots, \delta_{p,J-1})'$ and
$\gamma_{p}=(\gamma_{p,1}, \gamma_{p,2}, \allowbreak{\cdots,
\gamma_{p,J-1}})'$ where $\gamma_{p,1} = 0$ and recall that the first
cut-point $\gamma_{p,0} = -\infty$ and the last cut-point $\gamma_{p,J} =
\infty$.

The transformed cut-point vector $\delta_{p}$ does not have a known
conditional distribution and is sampled using an MH algorithm with a
random-walk proposal density. A tailored MH algorithm, as done in
\citet{Jeliazkov-etal-2008}, was attempted but later aborted because of the
increased computational time due to maximization of the likelihood function
at each iteration. However, estimates obtained from both forms of the MH
algorithm are identical.

Once the difficulties related to the cut-points have been addressed, the
joint posterior distribution can be derived using the Bayes' theorem. The
joint posterior distribution for $\beta_{p}, \delta_{p}$, latent weight $w$
and latent data $z$, assuming the following independent normal priors,
\begin{align*}
\beta_{p}   & \sim   N(\beta_{p0}, B_{p0}), \hspace{0.5in}\mathrm{and}\\
\delta_{p}  & \sim   N(\delta_{p0}, D_{p0}),
\end{align*}
can be written as proportional to the product of the likelihood and the
priors as,
\begin{align}
\pi(z,\beta_{p}, \delta_{p},w|y)
            & \propto  f(y|z,\beta_{p},\delta_{p},w) \; \pi( z,\beta_{p},\delta_{p} | w)  \; \pi(w)  \notag \\
            & \propto  f(y|z,\beta_{p},\delta_{p},w) \; \pi(z | \beta_{p}, w)
              \; \pi(\beta_{p}, \delta_{p}) \; \pi(w)  \label{eq:JointPostORI}\\
            & \propto \Big\{ \prod_{i=1}^{n} f(y_{i}|z_{i},\delta_{p}) \Big\} \; \pi(z | \beta_{p},w)
              \; \pi(w) \; \pi(\beta_{p}) \; \pi(\delta_{p}), \notag
\end{align}
where the likelihood, based on $AL(0,1,p)$, uses the fact that given the
latent variable $z$ and the cut-points $\delta_{p}$, the observed $y_{i}$ is
independent of $\beta_{p}$ because \eqref{eq:cutpoints} determines $y_{i}$
given $(z_{i}, \delta_{p})$ with probability one and that relation is not
dependent on $\beta_{p}$. In the second line, the density $\pi(z |
\beta_{p},w)$ can be obtained from \eqref{eq:model3} and is given by $\pi(z |
\beta_{p},w) = \prod_{i=1}^{n} N(z_{i}|x'_{i}\beta_{p} + \theta w_{i},
\tau^{2} w_{i})$. The last line in equation \eqref{eq:JointPostORI} uses
prior independence between $\beta_{p}$ and $\delta_{p}$. With the help of
preceding explanations, the ``complete data'' posterior in equation
\eqref{eq:JointPostORI} can be written as,
\begin{align}
\pi(z,\beta_{p}, \delta_{p},w|y)
          & \propto  \bigg\{ \prod_{i=1}^{n} 1\{\gamma_{p,y_{i}-1} < z_{i} \leq
            \gamma_{p,y_{i}} \} \; N(z_{i}|x'_{i}\beta_{p} + \theta w_{i}, \tau^{2} w_{i})
            \; \mathcal{E}(w_{i}|1) \bigg\} \notag \\
          & \qquad \times  N(\beta_{p0}, B_{p0}) \; N(\delta_{p0}, D_{p0}).
          \label{eq:CompPostORI}
\end{align}
Using equation \eqref{eq:CompPostORI} and two identification constraints,
$\gamma_{p,1}=0$ and $V(\epsilon) = \frac{1-2p+2p^{2}} {p^{2}(1-p)^{2}}$
(fixed for a given $p$), the objects of interest $(z,\beta_{p}, \delta_{p},
w)$ can be sampled as presented in Algorithm~\ref{alg:algorithm1}.


\begin{table*}[!t]
\begin{algorithm}[Sampling in $\mathrm{OR_{I}}$ model]
\label{alg:algorithm1} \rule{\textwidth}{0.5pt} \small{
\begin{itemize}
\item    Sample $\beta_{p}| z,w$ $\sim$  $N(\tilde{\beta}_{p}, \tilde{B}_{p})$, where,
\item[]  $\tilde{B}^{-1}_{p} = \bigg(\sum_{i=1}^{n}
        \frac{x_{i} x'_{i}}{\tau^{2} w_{i}} + B_{p0}^{-1} \bigg) $ \hspace{0.05in}
         and \hspace{0.05in} $\tilde{\beta}_{p} = \tilde{B}_{p}\bigg( \sum_{i=1}^{n}
         \frac{x_{i}(z_{i} - \theta w_{i})}{\tau^{2} w_{i}} + B_{p0}^{-1} \beta_{p0}
         \bigg)$.
\item    Sample $w_{i}|\beta_{p}, z_{i}$ $\sim$  $GIG \, (0.5, \tilde{\lambda}_{i},
    \tilde{\eta}) $, for $i=1,\cdots,n$, where,
\item[]  $\tilde{\lambda}_{i} = \Big( \frac{ z_{i} - x'_{i}\beta_{p}}{\tau}
    \Big)^{2}$ \hspace{0.05in} and \hspace{0.05in} $\tilde{\eta} = \Big(
    \frac{\theta^{2}}{\tau^{2}} + 2 \Big)$.
\item    Sample $\delta_{p}|y,\beta_{p}$ marginally of $w$ (latent weight) and $z$
    (latent data), by generating $\delta_{p}'$ using a random-walk chain $\delta'_{p}
    = \delta_{p} + u$, where $u \sim N(0_{J-2}, \iota^{2} \hat{D})$, $\iota$ is a tuning
    parameter and $\hat{D}$ denotes negative inverse Hessian, obtained by maximizing
     the log-likelihood with respect to $\delta_{p}$. Given the current value
     of $\delta_{p}$ and the proposed draw $\delta'_{p}$, return $\delta'_{p}$
     with probability,
         \begin{equation*}
          \alpha_{MH}(\delta_{p}, \delta'_{p}) = \min \bigg\{1,
          \frac{ \;f(y|\beta_{p},\delta'_{p}) \;\pi(\beta_{p}, \delta'_{p})}
          {f(y|\beta_{p},\delta_{p}) \;\pi(\beta_{p}, \delta_{p})}
          \bigg\};
         \end{equation*}
         otherwise repeat the old value $\delta_{p}$. The variance of $u$ may be
         tuned as needed for appropriate step size and acceptance rate.
\item    Sample $z_{i}|y, \beta_{p},\gamma_{p},w$ $\sim$ $TN_{(\gamma_{p,j-1},
    \gamma_{p,j})}(x'_{i}\beta_{p} + \theta w_{i}, \tau^{2}w_{i})$ for $i=1,\cdots,n$,
    where $\gamma_{p}$ is obtained by one-to-one mapping between $\gamma_{p}$ and
    $\delta_{p}$ from equation \eqref{eq:logtransformation}.
\end{itemize}}
\rule{\textwidth}{0.5pt}
\end{algorithm}
\end{table*}


The sampling algorithm for $\mathrm{OR_{I}}$ model is fairly straightforward
and primarily involves drawing parameters, with the exception of
$\delta_{p}$, from their conditional distributions. The parameter $\beta_{p}$
conditional on $z$ and $w$ follows a normal distribution, draws from which
are a routine exercise in econometrics. On the other hand, the conditional
distribution of latent weight $w$ follows a Generalized Inverse Gaussian
(GIG) distribution, draws from which can be obtained either by the ratio of
uniforms method \citep{Dagpunar-1988, Dagpunar-1989} or the envelope
rejection method \citep{Dagpunar-2007}. The transformed cut-point vector
$\delta_{p}$, as mentioned earlier, does not have a known conditional
distribution and is sampled using the MH algorithm, marginally of $(z, w)$
because the full likelihood (\ref{eq:likelihood}) conditional on $(\beta_{p},
\delta_{p})$ is independent of $(z, w)$. Sampling of cut-points from the full
likelihood was also employed in \citet{Jeliazkov-etal-2008}. Finally, the
latent variable $z$ conditional on $(y, \beta_{p}, \gamma_{p}, w)$ is sampled
from a truncated normal distribution, where the region of truncation is
determined based on a one-to-one mapping from $\delta_{p}$ using equation
\eqref{eq:logtransformation}. The derivations of the full conditional
distributions of ($\beta_{p}, w, z$) and details on the MH sampling of the
cut-point vector $\delta_{p}$ are presented in \ref{sec:appendixORI}.

The $\mathrm{OR_{I}}$ model can also be estimated using an alternative
identification scheme where the scale restriction is enforced by fixing a
second cut-point, for example $\gamma_{p,2}=1$. Fixing a second cut-point
would introduce the scale parameter $\sigma_{p}$ into the model
\eqref{eq:model3} and consequently, add another sampling block in
Algorithm~\ref{alg:algorithm1}. This identification scheme, although
plausible, seems unnecessary for the $\mathrm{OR_{I}}$ model and may lead to
higher inefficiency factors; however, it can be gainfully utilized when
number of outcomes equals 3 and is described in Section~\ref{sec:ORII}.

\subsubsection{Simulation Study}\label{sec:ORIsimStudy}

A simulation study was carried out to examine the performance of the
algorithm proposed in Section~\ref{sec:ORIest} and compare model fit with the
ordinal probit model (see Algorithm~2 in \citet{Jeliazkov-etal-2008}). In
particular, 300 observations were generated from the model $z_{i} = x'_{i}
\beta + \epsilon_{i}$, with $\beta$ = $(-2 \; 3 \; 4)'$, covariates were
generated from standard uniform distributions, and $\epsilon$ was generated
from a mixture of logistic distributions, $\mathcal{L}(-5,\pi^{2}/3)$ and
$\mathcal{L}(2,\pi^{2}/3)$, with mix proportions 0.3 and 0.7, respectively.
The histogram of the continuous variable $z$ (not shown) was approximately
unimodal and negatively skewed. The discrete response variable $y$ was
constructed based on the cut-point vector $\gamma = (0, 2, 3)$. In the
simulated data, the number of observations corresponding to the four
categories of $y$ were 107, 43, 36 and 114, respectively.

\begin{table}[!b]
\centering \footnotesize \setlength{\tabcolsep}{3pt} \setlength{\extrarowheight}{2pt}
\setlength\arrayrulewidth{1pt}
\begin{tabular}{c rrr rrr rrr rrr rrrr r  }
\toprule
& & \multicolumn{3}{c}{\textsc{25th quantile}} & & \multicolumn{3}{c}{\textsc{50th quantile}} & &   \multicolumn{3}{c}{\textsc{75th quantile}}
           & &   \multicolumn{3}{c}{\textsc{ord probit}}                \\
\cmidrule{3-5} \cmidrule{7-9}  \cmidrule{11-13} \cmidrule{15-17}
\textsc{parameters}  & &  \textsc{mean} & \textsc{std} & \textsc{if}
                     & &  \textsc{mean} & \textsc{std} & \textsc{if}
                     & &  \textsc{mean} & \textsc{std} & \textsc{if}
                     & &  \textsc{mean} & \textsc{std} & \textsc{if} &  \\
\midrule
$\beta_{1}$     & & $-1.62$  & $0.39$  & $2.25$  & & $-0.75$ & $0.31$ & $1.95$
                & & $ 0.10$  & $0.36$  & $2.27$  & & $-0.70$ & $0.18$ & $1.22$ \\
$\beta_{2}$     & & $ 1.20$  & $0.50$  & $2.13$  & & $ 1.64$ & $0.42$ & $2.02$
                & & $ 2.44$  & $0.49$  & $2.66$  & & $ 1.08$ & $0.22$ & $1.21$ \\
$\beta_{3}$     & & $ 1.12$  & $0.51$  & $2.00$  & & $ 1.96$ & $0.45$ & $2.29$
                & & $ 2.97$  & $0.50$  & $2.58$  & & $ 1.22$ & $0.23$ & $1.23$\\
$\delta_{1}$    & & $ 0.01$  & $0.15$  & $2.36$  & & $-0.22$ & $0.15$ & $3.11$
                & & $ 0.23$  & $0.15$  & $4.54$  & & $-0.86$ & $0.13$ & $2.66$ \\
$\delta_{2}$    & & $ 0.07$  & $0.16$  & $2.19$  & & $-0.34$ & $0.15$ & $2.59$
                & & $-0.09$  & $0.16$  & $3.43$  & & $-0.97$ & $0.14$ & $2.33$ \\
\bottomrule
\end{tabular}
\caption{Posterior mean (\textsc{mean}), standard
deviation (\textsc{std}) and inefficiency factor (\textsc{if}) of the parameters in the
25th, 50th and 75th quantile ordinal models and ordinal probit model.}
\label{Table:SimResult1}
\end{table}


The posterior estimates for the model parameters were obtained based on the
simulated data and the following independent normal priors: $\beta_{p} \sim
N(0_{3}, I_{3})$ and $\delta_{p} \sim N(0_{2}, 0.25 \,I_{2})$, for $p=(0.25,
0.5, 0.75)$. The same priors were utilized in the estimation of the ordinal
probit model. Use of less informative priors only causes a minor change in
the posterior estimates. Table~\ref{Table:SimResult1} reports the MCMC
results obtained from 12,000 iterations, after a burn-in of 3,000 iterations,
along with the inefficiency factors calculated using the batch-means method
(see \citealp{Greenberg-2012}). MH acceptance rate for $\delta_{p}$ was
around 30\% for all values of $p$ and $\iota=\sqrt{3}$. Convergence of MCMC
draws, as observed from the trace plots (not shown), was quick and occurred
within a few hundred iterations. The sampling took approximately 120 and 50
seconds per $1,000$ iterations for the quantile and ordinal probit models,
respectively.

The quantile ordinal models offer several choice of quantiles and one may
interpret the various choices of $p$ as corresponding to different family of
link functions. In such a scenario, model selection criterion such as
deviance information criterion or DIC
\citep{Spiegelhalter-etal-2002,{Celeux-etal-2006}} may be utilized to choose
a value of $p$ that is most consistent with the data. To illustrate, DIC was
computed for the 25th, 50th and 75th quantile models and the numbers were
$755.67$, $721.81$ and $704.16$, respectively. The DIC for the ordinal probit
model was $721.56$. Hence, amongst all the models considered, the 75th
quantile model provides the best fit, which is correct since the distribution
of the continuous variable $z$ is negatively skewed and so is the AL
distribution for $p=0.75$. The median quantile model and ordinal probit model
give almost identical DICs, but both provide a poorer fit compared to the
75th quantile model.

\subsection{$\mathrm{OR_{II}}$ Model} \label{sec:ORII}

The term ``$\mathrm{OR_{II}}$ model,'' as used in the paper, refers to an
ordinal model in which the number of outcomes equals three ($J = 3$), and
both location and scale restrictions are achieved through fixing cut-points.
In this case, fixing a second cut-point simplifies the sampling procedure,
since with three outcomes there are only two cut-points, both of which are
known by virtue of being fixed, so $\gamma_{p} = \gamma$ for all $p$.
However, use of this identification scheme introduces a scale parameter in
the model \eqref{eq:model3} that needs to be estimated. The
$\mathrm{OR_{II}}$ model can be written as follows,
\begin{equation}
\begin{split}
& z_{i}   = x'_{i} \beta_{p}  + \sigma_{p} \epsilon_{i}
          = x'_{i} \beta_{p}  + \sigma_{p} \theta w_{i} + \sigma_{p} \tau \sqrt{w_{i}} \,u_{i},
             \hspace{1.10in} \forall \; i=1, \cdots, n, \\
& \gamma_{j-1} < z_{i} \leq \gamma_{j}  \; \Rightarrow \; \emph{$y_{i}$ = j},
             \hspace{1.7 in} \forall \; i=1,\cdots, n; \; j=1,2, 3,
\end{split}
\label{eq:ORII}
\end{equation}
where $\sigma_{p}$ is the scale parameter at quantile $p$ and ($\gamma_{1},
\gamma_{2}$) are fixed at some values, in addition to $\gamma_{0}=-\infty$
and $\gamma_{3}=\infty$. Note that the scale parameter will be dependent on
$p$ because it will adjust to capture the variability in the data, since
$V(\epsilon)$ is constant for a given $p$.

\subsubsection{Estimation} \label{sec:ORIIest}

Estimation of the $\mathrm{OR_{II}}$ model \eqref{eq:ORII}, although free of
the MH algorithm, cannot directly utilize Gibbs sampling since the
conditional mean of $z_{i}| \beta_{p}, w_{i}$ involves the scale parameter
$\sigma_{p}$ \citep{Kozumi-Kobayashi-2011}. However, the scale parameter
$\sigma_{p}$ can be removed from the conditional mean through a simple
reformulation as follows,
\begin{equation}
z_{i} = x'_{i}\beta_{p} + \theta \nu_{i} + \tau \sqrt{\sigma_{p} \nu_{i}}\, u_{i},
\label{eq:model4}
\end{equation}
where $\nu_{i} = \sigma_{p} w_{i}$ and consequently,
$z_{i}|\beta_{p},\sigma_{p},\nu_{i} \sim N(x'_{i}\beta_{p} + \theta \nu_{i},
\tau^{2} \sigma_{p} \nu_{i})$. In the current formulation \eqref{eq:model4},
the $\mathrm{OR_{II}}$ model becomes conducive to Gibbs sampling. The next
step relates to prior distributions and they were specified as,
\begin{align*}
\beta_{p}       & \sim     N(\beta_{p0}, B_{p0}),\\
\sigma_{p}      & \sim     IG(n_{0}/2, d_{0}/2),\\
\nu_{i}         & \sim     \mathcal{E}(\sigma_{p}),
\end{align*}
where $IG$ and $\mathcal{E}$ stand for inverse-gamma and exponential
distributions, respectively. Employing Bayes' theorem, the joint posterior
distribution for ($z, \beta_{p}, \nu, \sigma_{p}$) can be written as
proportional to the product of the likelihood and the priors,
\begin{equation}
\begin{split}
\pi(z,\beta_{p}, \nu, \sigma_{p} | y)
        & \propto
         f(y|z, \beta_{p}, \nu, \sigma_{p} ) \; \pi(z|\beta_{p}, \nu, \sigma_{p} ) \;
        \pi(\nu|\sigma_{p}) \; \pi(\beta_{p}) \; \pi(\sigma_{p}),  \\
        & \propto
        \Big\{ \prod_{i=1}^{n} f(y_{i}|z_{i}, \sigma_{p})  \Big\} \; \pi(z|\beta_{p}, \nu, \sigma_{p} ) \;
        \pi(\nu|\sigma_{p}) \; \pi(\beta_{p}) \; \pi(\sigma_{p}),
\end{split}
\label{eq:JointPostORII}
\end{equation}
where the likelihood uses the property that $\sigma_{p} \, \epsilon \sim
AL(0, \sigma_{p}, p)$ and given the known cut-points and latent data $z$, the
observed $y_{i}$ does not depend on ($\beta_{p}, \nu$). The conditional
distribution of the latent data $z$ can be obtained from \eqref{eq:model4} as
$\pi(z|\beta_{p},\sigma_{p},\nu) = \prod_{i=1}^{n} N(x'_{i}\beta_{p} + \theta
\nu_{i}, \tau^{2} \sigma_{p} \nu_{i})$. Combining the likelihood, conditional
distribution of $z$ and the priors, the ``complete data'' posterior in
\eqref{eq:JointPostORII} can be expressed as,
\begin{align}
  \pi(z,\beta_{p}, \nu, \sigma_{p} | y)
     &  \propto
        \bigg\{ \prod_{i=1}^{n} 1\{\gamma_{y_{i}-1} < z_{i} \leq \gamma_{y_{i}} \} \;
        N(z_{i}|x'_{i}\beta_{p} + \theta \nu_{i},\tau^{2} \sigma_{p} \nu_{i}) \;
        \mathcal{E}(\nu_{i}|\sigma_{p}) \bigg\} \notag\\
     & \qquad \times N(\beta_{p0}, B_{p0}) \; IG(n_{0}/2, d_{0}/2), \label{eq:CompPostORII}
\end{align}
which can be utilized to derive the full conditional distributions for all
parameters of interest namely, $\beta_{p}$, $\nu$ and $\sigma_{p}$. The
derivations, presented in \ref{sec:appendixORII}, require collecting terms
for a parameter of interest assuming other parameters are known and then
identifying the distribution for the parameter of interest. Following this
intuitively simple approach, the parameters can be sampled from their
conditional posteriors as presented in Algorithm~\ref{alg:algorithm2}.

\begin{table*}[!t]
\begin{algorithm}[Sampling in $\mathrm{OR_{II}}$ model]
\label{alg:algorithm2} \rule{\textwidth}{0.5pt} \small{
\begin{itemize}
\item    Sample $\beta_{p}| z,\sigma_{p},\nu$ $\sim$
        $N(\tilde{\beta}_{p}, \tilde{B}_{p})$,  where,
\item[]  $\tilde{B}^{-1}_{p} = \bigg(\sum_{i=1}^{n} \frac{x_{i} x'_{i}}{\tau^{2}
    \sigma_{p} \nu_{i}} + B_{p0}^{-1} \bigg) $   \hspace{0.05in} and \hspace{0.05in}
    $\tilde{\beta}_{p} = \tilde{B}_{p}\bigg( \sum_{i=1}^{n} \frac{x_{i}(z_{i} - \theta
    \nu_{i})}{\tau^{2} \sigma_{p} \nu_{i}} + B_{p0}^{-1} \beta_{p0} \bigg)$.
\item    Sample $\sigma_{p}| z, \beta_{p}, \nu$ $\sim$ $IG(\tilde{n}/2,
    \tilde{d}/2)$, where,
\item[]  $\tilde{n} = (n_{0} + 3n)$ \hspace{0.05in} and \hspace{0.05in} $\tilde{d} =
    \sum_{i=1}^{n}(z_{i} - x'_{i}\beta_{p} - \theta \nu_{i})^{2}/\tau^{2}\nu_{i} + d_{0}
    + 2 \sum_{i=1}^{n} \nu_{i}$.
\item    Sample $\nu_{i}| z_{i}, \beta_{p}, \sigma_{p}$ $\sim$ $GIG(0.5,
    \tilde{\lambda_{i}}, \tilde{\eta})$, for $i=1,\cdots,n$, where,
\item[]  $\tilde{\lambda_{i}}= \frac{ ( z_{i} - x'_{i}\beta_{p} )^2
    }{\tau^{2}\sigma_{p}}$ \hspace{0.05in} and \hspace{0.05in} $\tilde{\eta} = \Big(
    \frac{\theta^2}{\tau^{2} \sigma_{p}} + \frac{2}{\sigma_{p}} \Big)$.
\item    Sample $z_{i}|y, \beta_{p},\sigma_{p}, \nu_{i}$ $\sim$ $TN_{(\gamma_{j-1},
    \gamma_{j})}(x'_{i}\beta_{p} + \theta \nu_{i}, \tau^{2} \sigma_{p} \nu_{i})$ for
    $i=1,\cdots,n$, and $j=1,2,3$.
\end{itemize}}
\rule{\textwidth}{0.5pt}
\end{algorithm}
\end{table*}


\subsubsection{Simulation Study}\label{sec:ORIIsimStudy}

A simulation study was carried out to examine the performance of the
algorithm proposed in Section~\ref{sec:ORIIest} and compare model fit with
the ordinal probit model (see Algorithm~3 in \citet{Jeliazkov-etal-2008}). In
particular, 300 observations were generated from the model $z_{i} =
x'_{i}\beta + \epsilon_{i}$, with $\beta$ = $(2 \; 2 \; 1)'$, covariates were
generated from a standard bivariate normal distribution with correlation 0.25
and $\epsilon$ was generated from a mixture of Gaussian distributions, $N(-6,
4)$ and $N(5,1)$, with mix proportions 0.3 and 0.7, respectively. The
histogram of continuous variable $z$ (not shown) was bimodal and negatively
skewed. The discrete response variable $y$ was constructed based on cut-point
vector $\gamma$ = ($0$, $4$). In the simulated data, the number of
observations corresponding to the three category of $y$ were 77, 38 and 185,
respectively.

The posterior estimates in the quantile ordinal models were obtained based on
the following priors: $\beta_{p} \sim N(0_{3}, I_{3})$, $\sigma_{p} \sim
IG(5/2, 8/2)$ and $\nu \sim \mathcal{E}(\sigma_{p})$ for $p=(0.25,0.5,0.75)$.
Posterior estimation of ordinal probit model used the same prior on $\beta$
and $\sigma^2 \sim IG(5/2, 8/2)$. The hyperparameters of the inverse-gamma
distribution was chosen to keep the prior less informative. For all the
models, the cut-points were fixed at ($0$, $4$), which are same as that used
to construct the discrete response $y$. Table~\ref{Table:SimResult2} reports
the Gibbs sampling results obtained from 12,000 iterations after a burn-in of
3,000 iterations, along with the inefficiency factors calculated using the
batch-means method (see \citealp{Greenberg-2012}). Convergence of MCMC draws,
as observed from the trace plots (not shown), occurred within a few hundred
iterations. The sampling procedure took approximately 87 and 75 seconds per
1,000 iterations for the quantile and ordinal probit models, respectively.

\begin{table}[!t]
\centering \footnotesize \setlength{\tabcolsep}{3pt} \setlength{\extrarowheight}{2pt}
\setlength\arrayrulewidth{1pt}
\begin{tabular}{c rrr rrr rrr rrr rrr r  }
\toprule
& & \multicolumn{3}{c}{\textsc{25th quantile}}
& & \multicolumn{3}{c}{\textsc{50th quantile}}
& & \multicolumn{3}{c}{\textsc{75th quantile}}
& & \multicolumn{3}{c}{\textsc{ord probit}}   \\
\cmidrule{3-5} \cmidrule{7-9}  \cmidrule{11-13} \cmidrule{15-17}
\textsc{parameters}  & &  \textsc{mean} & \textsc{std} & \textsc{if}
                     & & \textsc{mean} & \textsc{std} & \textsc{if}
                     & & \textsc{mean} & \textsc{std} & \textsc{if}
                     & & \textsc{mean} & \textsc{std} & \textsc{if}  \\
\midrule
$\beta_{1}$   & & $ 0.27$  & $0.58$  & $2.12$  & & $5.03$ & $0.41$ & $2.31$
              & & $ 7.05$  & $0.37$  & $5.47$  & & $4.62$ & $0.49$ & $1.28$  \\
$\beta_{2}$   & & $ 1.70$  & $0.57$  & $1.96$  & & $1.63$ & $0.39$ & $2.10$
              & & $ 1.67$  & $0.31$  & $4.50$  & & $2.05$ & $0.49$ & $1.29$  \\
$\beta_{3}$   & & $ 0.41$  & $0.57$  & $1.88$  & & $0.64$ & $0.41$ & $2.04$
              & & $ 0.70$  & $0.30$  & $3.14$  & & $0.68$ & $0.50$ & $1.21$  \\
$\sigma$      & & $ 4.03$  & $0.60$  & $4.42$  & & $3.28$ & $0.43$ & $3.95$
              & & $ 1.41$  & $0.15$  & $4.11$  & & $7.79$ & $0.95$ &  $3.41$ \\
\bottomrule
\end{tabular}
\caption{Posterior mean (\textsc{mean}), standard
deviation (\textsc{std}) and inefficiency factor (\textsc{if}) of the parameters in the
25th, 50th and 75th quantile ordinal models and ordinal probit model.}
\label{Table:SimResult2}
\end{table}


In order to compare the 25th, 50th, 75th quantile models and the ordinal
probit model, the DIC was computed and the values were $540.61$, $528.96$,
$526.19$ and $533.80$, respectively. Therefore, amongst the quantile models,
the 75th quantile model provides the best fit, which is correct since the
distribution of the continuous variable $z$ is negatively skewed and so is
the AL distribution for $p=0.75$. In addition, both the 50th and 75th
quantile models provide a better fit than the ordinal probit model.

\section{Application}\label{sec:applications}

In this section, the proposed algorithms for quantile estimation of ordinal
models are used in two applications that are of interest in economics and the
broader social sciences. The first application uses the National Longitudinal
Study of Youth (NLSY, 1979) survey data from \citet{Jeliazkov-etal-2008} to
analyze the topic of educational attainment and extends the analysis to the
domain of quantile regression. The second application uses the American
National Election Studies (ANES) survey data to evaluate public opinion on
raising federal income taxes for individuals' who make more than \$250,000
per year.

\subsection{Educational Attainment}

In this application, the NLSY data taken from \citet{Jeliazkov-etal-2008} was
utilized to study educational attainment. The NLSY was started in 1979 with
more than 12,000 youths to conduct annual interviews on a wide range of
demographic questions. However, the sample used by
\citet{Jeliazkov-etal-2008} contains data on 3923 individuals only, because
the analysis was restricted to cohorts aged 14-17 in 1979 and for whom family
income variable could be constructed.


\begin{table}[!b]
\centering \footnotesize \setlength{\tabcolsep}{3pt} \setlength{\extrarowheight}{2pt}
\setlength\arrayrulewidth{1pt}
\begin{tabular}{l rrc rrc rrc rrc }
\toprule
  & & \multicolumn{2}{c}{\textsc{25th quantile}} & &   \multicolumn{2}{c}{\textsc{50th quantile}}
  & & \multicolumn{2}{c}{\textsc{75th quantile}} & & \multicolumn{2}{c}{\textsc{ord probit}}          \\
  \cmidrule{3-4} \cmidrule{6-7}  \cmidrule{9-10} \cmidrule{12-13}
\textsc{parameters} & &  \textsc{mean} & \textsc{std}  & & \textsc{mean} & \textsc{std}
                    & &  \textsc{mean} & \textsc{std}  & &  \textsc{mean} & \textsc{std} \\
\midrule
\textsc{intercept}
& &  $-5.92$  & $0.33$   & &  $-3.18$  & $0.22$  & &  $-0.61$  & $0.27$
                         & &  $-1.34$  & $0.09$  \\
\textsc{family income (sq. rt.)}
& &  $ 0.39$  & $0.04$   & &  $ 0.28$  & $0.02$  & &  $ 0.28$  & $0.03$
                         & &  $ 0.14$  & $0.01$  \\
\textsc{mother's education}
& &  $ 0.18$  & $0.03$   & &  $ 0.12$  & $0.02$  & &  $ 0.12$  & $0.02$
                         & &  $ 0.05$  & $0.01$  \\
\textsc{father's education}
& &  $ 0.21$  & $0.02$   & & $ 0.18$   & $0.02$  & &  $ 0.17$  & $0.02$
                         & & $ 0.07$   & $0.01$  \\
\textsc{mother worked}
& &  $ 0.08$  & $0.10$   & & $ 0.07$   & $0.08$  & &  $ 0.06$  & $0.10$
                         & & $ 0.03$   & $0.04$  \\
\textsc{female}
& &  $ 0.58$  & $0.10$   & & $ 0.35$   & $0.08$  & &  $ 0.23$  & $0.09$
                         & & $ 0.16$   & $0.04$  \\
\textsc{black}
& &  $ 0.64$  & $0.13$   & & $ 0.43$   & $0.09$  & &  $ 0.25$  & $0.11$
                         & & $ 0.15$   & $0.04$  \\
\textsc{urban}
& &  $-0.42$  & $0.14$   & & $-0.08$   & $0.09$  & &  $ 0.13$  & $0.11$
                         & & $-0.05$   & $0.04$  \\
\textsc{south}
& &  $ 0.13$  & $0.13$   & & $ 0.08$   & $0.08$  & &  $ 0.15$  & $0.10$
                         & & $ 0.05$   & $0.04$  \\
\textsc{age cohort 2}
& &  $-0.09$  & $0.23$   & & $-0.05$   & $0.12$  & &  $-0.03$  & $0.14$
                         & & $-0.03$   & $0.05$  \\
\textsc{age cohort 3}
& &  $-0.06$  & $0.16$   & & $-0.05$   & $0.12$  & &  $ 0.04$  & $0.15$
                         & & $ 0.00$   & $0.05$  \\
\textsc{age cohort 4}
& &  $ 0.50$  & $0.16$   & & $ 0.49$   & $0.13$  & &  $ 0.54$  & $0.15$
                         & & $ 0.23$   & $0.06$  \\
\quad $\delta_{1}$
& &  $ 1.11$  & $0.03$   & & $ 0.90$   & $0.03$  & &  $ 1.27$  & $0.03$
                         & & $ 0.08$   & $0.02$  \\
\quad $\delta_{2}$
& &  $ 1.13$  & $0.03$   & & $ 0.55$   & $0.03$  & &  $ 0.56$  & $0.03$
                         & & $-0.28$   & $0.03$      \\
\bottomrule
\end{tabular}
\caption{Posterior mean (\textsc{mean}) and standard
deviation (\textsc{std}) of model parameters in the educational attainment application.
Identification achieved through variance restriction.}
\label{Table:EducAppEst}
\end{table}


The dependent variable in the model, education degrees, has four categories:
\emph{(i) less than high school}, \emph{(ii) high school degree}, \emph{(iii)
some college or associate's degree}, and \emph{(iv) college or graduate
degree}, and the number of observations corresponding to each category are
897 (22.87\%), 1392 (35.48\%), 876 (22.33\%) and 758 (19.32\%), respectively.
The independent variables included in the model are as follows: square root
of family income, mother's education, father's education, mother's working
status, gender, race, and whether the youth lived in an urban area or the
South at the age of 14. In addition, to control for age cohort effects, three
indicator variables were included to indicate an individuals' age in 1979.
Using data on the above variables, the application studies the effect of
family background, individual and school variables on educational attainment.

In this application, there are four outcomes and hence the fixed variance
restriction was utilized to estimate the quantile models (using the
$\mathrm{OR_{I}}$ framework) and the ordinal probit model. Priors on the
parameters were same as that in the simulation study of
Section~\ref{sec:ORIsimStudy}. Table~\ref{Table:EducAppEst} reports the
results obtained from 12,000 iterations after a burn-in of 3,000 iterations.
Inefficiency factors, not reported, were less than 6 for all the parameters
and MCMC simulation draws converged to the target distribution within a few
hundred iterations.

The results for the ordinal probit model ($\epsilon \sim N(0,I)$), presented
in the last two columns of Table~\ref{Table:EducAppEst}, are identical to
that obtained by \citet{Jeliazkov-etal-2008}. It is seen that the signs of
the coefficients are mostly consistent with what is typically found in the
literature. For example, parental education and higher family income have a
positive effect on educational attainment. Similarly, mother's labor force
participation has a positive effect on educational attainment. The table also
shows that conditional on other covariates, females, blacks or individuals'
from South have higher educational attainment, respective to the base
categories. On the other hand, living in an urban area has a negative effect
on educational attainment and the age cohort variables have a different
effect on the educational attainment of an individual.

To supplement the analysis, Table~\ref{Table:EducAppEst} also presents the
posterior estimates and standard deviations of the parameters for the ordinal
quantile regression model ($\epsilon \sim AL(0,1\allowbreak{,p)}$), estimated
for $p=0.25, 0.50$, and $0.75$. It is seen that the sign and magnitude of the
estimates for the quantile models are somewhat similar to that obtained for
the ordinal probit model. However, this does not imply similar inferences, as
explained in the next paragraph. The last two rows of
Table~\ref{Table:EducAppEst}, present the posterior mean and standard
deviation of the transformed cut-point vector $\delta_{p} = (\delta_{1,p},
\delta_{2,p})'$. Clearly, the estimated $\delta$ for the ordinal probit model
is different compared to the quantile regression models. This difference is
related to different distributional assumptions associated with the models.


\begin{table}[!b]
\centering \footnotesize \setlength{\tabcolsep}{3pt} \setlength{\extrarowheight}{4pt}
\setlength\arrayrulewidth{1pt}
\begin{tabular}{l rcc ccc cc}
\midrule
     & & \textsc{25th quantile} & & \textsc{50th quantile} & &  \textsc{75th quantile}
     & & \textsc{ord probit}    \\
                     \midrule
$\Delta$P(high school dropout)
& &  $-0.0415$            & &   $-0.0313$             & &  $-0.0193$
                          & &   $-0.0390$     \\
$\Delta$P(high school degree)
& &  $\phantom{+}0.0022$  & &   $-0.0133$             & &  $-0.0186$
                          & &   $-0.0000$     \\
$\Delta$P(college or associate's)
& &  $\phantom{+}0.0204$  & &   $\phantom{+}0.0201$   & &  $\phantom{+}0.0097$
                          & &   $\phantom{+}0.0000$   \\
$\Delta$P(college or graduate)
& &  $\phantom{+}0.0188$  & &   $\phantom{+}0.0246$   & &  $\phantom{+}0.0282$
                          & &   $\phantom{+}0.0390$   \\
\midrule
\end{tabular}
\caption{Change in predicted probabilities for a \$10,000
increase in income.}
\label{Table:EducAppCovariateEffect}
\end{table}


In addition to the above analysis, it is necessary to emphasize that the link
functions associated with ordinal probit and quantile ordinal models are
non-linear and non-monotonic, as such an interpretation of the resulting
parameter estimates is not straightforward since the coefficients by
themselves do not give the impact of a change in one or more of the
covariates. Consequently, the $\beta$ estimates from the quantile models and
ordinal probit model do not imply the same covariate effects. To explain and
highlight the differences in covariate effects, I computed the effect of a
\$10,000 increase in family income on educational outcomes, marginalized over
parameters and remaining covariates. The results are reported in
Table~\ref{Table:EducAppCovariateEffect}, which shows that the change in
predicted probabilities are different in ordinal probit and quantile models.
For example, the \$10,000 increase in income decreases the probability of
high school dropout by $0.039$ in the ordinal probit model. In contrast, the
same probability decreases by $0.0415$, $0.0313$ and $0.0193$ at the 25th,
50th and 75th quantiles, respectively.

Finally, an investigation on model selection based on the DIC reports the
following numbers: $9840.75$, $9781.02$, and $9977.30$ for the 25th, 50th and
75th quantile models, respectively. The DIC for the ordinal probit model was
$9736.21$. Hence, according to DIC the ordinal probit model provides the best
fit amongst the models considered, followed by the median model. However, it
may be possible that for another value of $p$, possibly close to $p=0.50$,
the DIC is lower than the DIC for the ordinal probit model.

\subsection{Tax Policy}

This application aims to analyze public opinion on a recently considered tax
policy: the proposal to raise federal income taxes for couples (individuals)
earning more than \$250,000 (\$200,000) per year (hereafter, termed
``pro-growth'' policy), and aims to identify factors that may increase or
decrease support in favor of the proposed policy.

The pro-growth policy, proposed by President Barack H. Obama in 2010, was
essentially aimed to extend the ``Bush Tax'' cuts for the lower and middle
income class, but restore higher rates for the richer class. It thereby aimed
to promote growth in the U.S. economy, struggling due to the recession, by
supporting consumption amongst the low-middle income families. The policy
became a subject of extended political debate with respect to the definition
of benchmark income, beneficiaries of the tax cuts and whether it would spur
sufficient growth. However, the proposed policy received a two-year extension
and was part of a larger tax and economic package, named the ``Tax Relief,
Unemployment Insurance Reauthorization, and Job Creation Act of 2010''. The
pro-growth policy re-surfaced in the 2012 presidential election and formed a
crucial point of discussion during the presidential debate.

\begin{table}[!b]
\centering \footnotesize \setlength{\tabcolsep}{6pt} \setlength{\extrarowheight}{2pt}
\setlength\arrayrulewidth{1pt}
\begin{tabular}{l lll rrr rrr r }
\toprule
\textsc{variables}       & &   \textsc{description} & \textsc{mean} & \textsc{count}                             \\
\midrule
\textsc{employed}        & & Indicator for individual being employed
                         & $0.53$   &  $ 621$ \\
\textsc{income}          & & Indicator for household income $> \$75,000$
                         & $0.30$   &  $ 346$       \\
\textsc{bachelors}       & & Individual's highest degree is Bachelors
                         & $0.20$   &  $ 235$           \\
\textsc{post-bachelors}  & & Highest degree is Masters, Professional or Doctorate
                         & $0.09$   &  $ 108$           \\
\textsc{computers}       & & Individual or household owns a computer
                         & $0.84$   &  $ 972$           \\
\textsc{cellphone}       & & Individual or household owns a cell phone
                         & $0.90$   &  $1,051$          \\
\textsc{white}           & & Race of the individual is white
                         & $0.86$   &  $1,004$          \\
\bottomrule
\end{tabular}
 \caption{Variable definitions and data summary.}
\label{Table:TaxAppVar}
\end{table}


The pro-growth policy was included as a survey question in the 2010-2012
American National Election Studies (ANES) on the Evaluations of Government
and Society Study 1 (EGSS 1). The ANES survey was conducted over the internet
using nationally representative probability samples and after removing
missing observations provides $1,164$ observations. The survey recorded
individuals' opinion as either \emph{oppose}, \emph{neither favor nor
oppose}, or \emph{favor} the tax increase. This forms the dependent variable
in the model with 263 (22.59\%), 261 (22.42\%) and 640 (54.98\%) observations
in the respective categories. In addition, the survey collected information
on a wide range of demographic variables, some of which were included as
independent variables in the model. They include employment status, income
level, education, computer ownership, cell phone ownership and race.
Definition of the variables are presented in Table~\ref{Table:TaxAppVar}
together with the mean and count on each of them.


\begin{table}[!b]
\centering \footnotesize \setlength{\tabcolsep}{3pt} \setlength{\extrarowheight}{2pt}
\setlength\arrayrulewidth{1pt}
\begin{tabular}{l rrr rrr rrr rrr}
\toprule
 & & \multicolumn{2}{c}{\textsc{25th quantile}}   & & \multicolumn{2}{c}{\textsc{50th quantile}}
 & & \multicolumn{2}{c}{\textsc{75th quantile}}   & & \multicolumn{2}{c}{\textsc{ord probit}}   \\
      \cmidrule{3-4} \cmidrule{6-7}  \cmidrule{9-10}   \cmidrule{12-13}
\textsc{parameters} & &  \textsc{mean} & \textsc{std} & &  \textsc{mean} & \textsc{std}
                    & & \textsc{mean} & \textsc{std}  & &  \textsc{mean} & \textsc{std}   \\
\midrule
\textsc{intercept}      & &  $ 1.00$ &  $0.46$  & & $ 2.10$  & $0.43$  & &  $ 3.42$  &  $0.37$  & &  $ 2.58$ &  $0.53$    \\
\textsc{employed}       & &  $-0.04$ &  $0.29$  & & $ 0.20$  & $0.26$  & &  $ 0.21$  &  $0.24$  & &  $ 0.16$ &  $0.32$   \\
\textsc{income}         & &  $-0.73$ &  $0.34$  & & $-0.46$  & $0.30$  & &  $-0.47$  &  $0.28$  & &  $-0.72$ &  $0.37$   \\
\textsc{bachelors}      & &  $-0.17$ &  $0.38$  & & $ 0.07$  & $0.33$  & &  $ 0.12$  &  $0.32$  & &  $-0.07$ &  $0.40$   \\
\textsc{post-bachelors} & &  $-0.02$ &  $0.44$  & & $ 0.43$  & $0.40$  & &  $ 0.53$  &  $0.39$  & &  $ 0.28$ &  $0.47$   \\
\textsc{computers}      & &  $ 0.02$ &  $0.35$  & & $ 0.62$  & $0.33$  & &  $ 0.61$  &  $0.29$  & &  $ 0.53$ &  $0.41$   \\
\textsc{cellphone}      & &  $ 0.38$ &  $0.42$  & & $ 0.78$  & $0.38$  & &  $ 0.75$  &  $0.32$  & &  $ 0.96$ &  $0.47$   \\
\textsc{white}          & &  $-0.82$ &  $0.36$  & & $ 0.02$  & $0.34$  & &  $ 0.29$  &  $0.30$  & &  $-0.29$ &  $0.41$   \\
\quad $\sigma$          & &  $ 1.96$ &  $0.12$  & & $ 1.99$  & $0.12$  & &  $ 1.00$  &  $0.06$  & &  $ 4.76$ &  $0.26$   \\
\bottomrule
\end{tabular}
\caption{Posterior mean (\textsc{mean}) and standard
deviation (\textsc{std}) of model parameters for the tax policy application.
Identification achieved through cut-point restrictions, i.e. $\gamma = (0, 3)$.}
\label{Table:TaxAppEst}
\end{table}


In this application, the dependent variable has three categories and hence
analyzed within the $\mathrm{OR_{II}}$ framework as presented in
Section~\ref{sec:ORIIest} (the application was also analyzed under the
$\mathrm{OR_{I}}$ framework and inferences were almost identical). The
ordinal probit model was also estimated for comparison purposes. Priors on
the parameters were same as that in the simulation study (see
Section~\ref{sec:ORIIsimStudy}) and estimates are based on 12,000 iterations
after a burn-in of 3,000 iterations. Inefficiency factors, not reported, were
less than 5 for all the parameters and MCMC draws converged to the target
distribution within a few hundred iterations.

Table~\ref{Table:TaxAppEst} reports the posterior estimates and standard
deviations of ($\beta_{p}, \sigma_{p}$) in the quantile models and ($\beta,
\sigma$) in the ordinal probit model. The $\beta$ coefficients point to some
interesting findings. For example, the indicator variable for income has a
negative effect on the probability of supporting the tax increase, which is
understandable since individuals' earning relatively higher income would like
to pay lower tax and would oppose the tax increase. In contrast, computer and
cell phone ownership indicators have a positive effect on the proposed tax
increase. This implies that access to information through ownership of
computers and cell phones, especially in the upper half of the distribution,
plays an important role in the decision to support the pro-growth policy and
highlights the significance of digital devices and the associated flow of
information on public opinion. Covariate effects calculated for income,
computer and cell phone ownerships show that in the ordinal probit model,
change in outcome probabilities are ($0.0473, 0.0131, -0.0604$), ($-0.0351,
-0.0092, 0.0442$) and ($-0.0653, -0.0150, 0.0808$), respectively. The
corresponding change in outcome probabilities for the quantile models are
presented in Table~\ref{Table:TaxAppCovEffect}, and it is seen that the
change in predicted probabilities are different at different quartiles for
all the variables.


\begin{table}[!t]\centering
\footnotesize \setlength{\tabcolsep}{2pt} \setlength{\extrarowheight}{2pt}
\begin{tabular}{lrr rrr rrr rrr r}
\toprule
    & &  \multicolumn{3}{c}{\textsc{income}} & & \multicolumn{3}{c}{\textsc{computer}} & & \multicolumn{3}{c}{\textsc{cell phone}} \\
    \cmidrule{3-5} \cmidrule{7-9}  \cmidrule{11-13}
    & &  25th   &  50th   & 75th    & &  25th     &  50th    &  75th
    & &  25th   &  50th   & 75th         \\ [0.5ex] \hline \\[-6pt]
$\Delta$P(oppose)
& &  $ 0.0630$   &  $ 0.0258$  & $ 0.0272$  & &  $-0.0019$   &  $-0.0356$  &  $-0.0360$  & &  $-0.0314$   &  $-0.0467$   & $-0.0443$    \\
$\Delta$P(neutral)
& &  $-0.0144$   &  $ 0.0285$  & $ 0.0304$  & &  $ 0.0015$   &  $-0.0375$  &  $-0.0401$  & &  $ 0.0093$   &  $-0.0454$   & $-0.0492$    \\
$\Delta$P(favor)
& &  $-0.0486$   &  $-0.0542$  & $-0.0576$  & &  $ 0.0004$   &  $ 0.0732$  &  $ 0.0761$  & &  $ 0.0250$   &  $ 0.0921$   & $ 0.0935$    \\
\bottomrule
\end{tabular}
\caption{Change in predicted probabilities as one goes from (a) less than
$\$75,000$ to more than $\$75,000$, (b) not owning a computer to owning a
computer, and (c) not owning a cell phone to owning a cell phone.}
\label{Table:TaxAppCovEffect}
\end{table}


To perform model comparison, DIC was calculated for the 25th, 50th and 75th
quantile ordinal models and the numbers were $2330.06$, $2336.73$ and
$2337.85$, respectively. The DIC for the ordinal probit model was $2335.89$.
Hence, according to DIC the 25th quantile model provides the best fit to the
data, followed by the ordinal probit model.

\section{Conclusion}\label{sec:conclusion}
The paper considers the Bayesian analysis of quantile regression models for
univariate ordinal data, and proposes a method that can be extensively
utilized in a wide class of applications across disciplines including
business, economics and social sciences. The method exploits the latent
variable inferential framework of \citet{Albert-Chib-1993} and capitalizes on
the normal-exponential mixture representation of the AL distribution.
Additionally, the scale restriction is judiciously chosen to simplify the
estimation procedure. In particular, when the number of outcomes $J$ is
greater than 3, attaining scale restriction by fixing the variance (termed
$\mathrm{OR_{I}}$ model) appears preferable. This is because fixing the
variance eliminates the need to sample the scale parameter $\sigma_{p}$.
Estimation utilizes a combination of Gibbs sampling and the MH algorithm
(only for transformed cut-points $\delta_{p}$). In the simplest case, when $J
= 3$ and scale restriction is realized by fixing a second cut-point (termed
$\mathrm{OR_{II}}$ model), the model does not have any unknown cut-points.
Consequently, the estimation of $\mathrm{OR_{II}}$ model relies solely on
Gibbs sampling.

The algorithms corresponding to $\mathrm{OR_{I}}$ and $\mathrm{OR_{II}}$
models are illustrated in Monte Carlo simulation studies with 300
observations, where the errors are generated from a mixture of logistic and
Gaussian distributions, respectively. Posterior means, standard deviations
and inefficiency factors are calculated for ($\beta_{p}, \delta_{p}$) in
$\mathrm{OR_{I}}$ model and ($\beta_{p}, \sigma_{p}$) in $\mathrm{OR_{II}}$
model. In both the models, posterior estimates of $\beta_{p}$ are
statistically different from zero, standard deviations are small and
inefficiency factors are all less than 6. The transformed cut-points
$\delta_{p}$ have an MH acceptance rate of around 30\% across quantiles for a
given value of the tuning parameter and inefficiency factors are all less
than 5. Similarly, inefficiency factors for the scale parameter $\sigma_{p}$
are all less than 5. Both the algorithms are reasonably fast and took
approximately 120 and 87 seconds per one thousand iterations, respectively.
Model comparison using DIC shows that the quantile ordinal models provide a
better model fit relative to the ordinal probit model.

The proposed techniques are applied to two studies in economics related to
educational attainment and public opinion on extending the ``Bust Tax'' cuts.
In the first application, the dependent variable, educational attainment, has
four categories and the $\mathrm{OR_{I}}$ framework is employed to estimate
the quantile ordinal models. Ordinal probit model is also estimated. It is
found that the sign of the estimated coefficients are mostly consistent with
what is typically found in the literature. In addition, the covariate effect
of a \$10,000 increase in income is shown to have a heterogeneous effect
across quantiles. Model comparison favors the ordinal probit model followed
by the median model. The second application analyzes the factors affecting
public opinion on raising federal income taxes for couples (individuals) who
make more than \$250,000 (\$200,000) per year. Opinions are classified into
three categories and  studied within the $\mathrm{OR_{II}}$ framework. It is
found that access to information through ownership of computers and cell
phones have a positive effect, but the income indicator variable (greater
than $\$75,000$) has a negative effect on the probability of supporting the
proposed tax increase and that these effects vary across quantiles. Model
comparison selects the 25th quantile model to be the best fitting model.


\newpage
\appendix
\renewcommand\thesection{Appendix \Alph{section}}


\section{Conditional Densities in $\mathrm{OR_{I}}$ Model}\label{sec:appendixORI}

In the $\mathrm{OR_{I}}$ model, the full conditional densities for
$\beta_{p}$, $w$ and latent variable $z$ are derived based on the complete
posterior density \eqref{eq:CompPostORI}. However, the transformed cut-point
vector $\delta_{p}$ does not have a tractable conditional distribution and is
sampled using the MH algorithm. The derivations below follow the ordering as
presented in Algorithm~\ref{alg:algorithm1}.

Starting with $\beta_{p}$, the full conditional density $\pi(\beta_{p}|z,w)$
is proportional to $\pi(\beta_{p}) \times \allowbreak{f(z|\beta_{p},w)}$ and
its kernel can be written as,
\begin{align*}
\pi(\beta_{p}|z,w)
          & \propto
          \exp\bigg[-\frac{1}{2} \bigg\{ \sum_{i=1}^{n}  \bigg( \frac{z_{i} - x'_{i}\beta_{p} - \theta w_{i}}
          {\tau \sqrt{w_{i}}} \bigg)^2  +
          (\beta_{p} - \beta_{p0} )' B_{p0}^{-1} (\beta_{p} - \beta_{p0} )   \bigg\}  \bigg]\\
          & \propto
          \exp\bigg[-\frac{1}{2} \bigg\{ \beta_{p}'
          \bigg( \sum_{i=1}^{n} \frac{x_{i} x'_{i}}{\tau^{2} w_{i}} + B_{p0}^{-1}\bigg) \beta_{p}
           - \beta'_{p} \bigg( \sum_{i=1}^{n} \frac{x_{i}(z_{i}-\theta w_{i})}{\tau^{2} w_{i}}
           + B_{p0}^{-1} \beta_{p0}  \bigg)     \\
          &  \hspace{35pt} - \bigg( \sum_{i=1}^{n} \frac{x'_{i}(z_{i}-\theta w_{i})}{\tau^{2} w_{i}}
           + \beta'_{p0}B_{p0}^{-1}  \bigg) \beta_{p}    \bigg\}          \bigg]\\
          & \propto
          \exp\bigg[ -\frac{1}{2} \bigg\{ \beta'_{p}\tilde{B}^{-1}_{p}\beta_{p}
          - \beta'_{p}\tilde{B}^{-1}_{p}\tilde{\beta}_{p} - \tilde{\beta}'_{p}
          \tilde{B}^{-1}_{p}\beta_{p}
          \bigg\}  \bigg],
\end{align*}
where the second line omits all terms not involving $\beta_{p}$ and the third
line introduces two terms, $\tilde{B}_{p}$ and $\tilde{\beta}_{p}$, which are
defined as follows,
\begin{equation*}
\tilde{B}^{-1}_{p} = \bigg(\sum_{i=1}^{n} \frac{x_{i} x'_{i}}{\tau^{2}
         w_{i}} + B_{p0}^{-1} \bigg)    \hspace{0.25in} \mathrm{and} \hspace{0.25in}
         \tilde{\beta}_{p} =
         \tilde{B}_{p}\bigg( \sum_{i=1}^{n} \frac{x_{i}(z_{i} - \theta w_{i})}{\tau^{2} w_{i}}
         + B_{p0}^{-1} \beta_{p0} \bigg).
\end{equation*}
Adding and subtracting $\tilde{\beta}'_{p}
\tilde{B}^{-1}_{p}\tilde{\beta}_{p}$ inside the curly braces, the square can
be completed as,
\begin{align*}
\pi(\beta_{p}|z,w)
          & \propto    \exp\bigg[ -\frac{1}{2}
          \bigg\{ \beta'_{p}\tilde{B}^{-1}_{p}\beta_{p}
          - \beta'_{p}\tilde{B}^{-1}_{p}\tilde{\beta}_{p}
          - \tilde{\beta}'_{p} \tilde{B}^{-1}_{p}\beta_{p}
          + \tilde{\beta}'_{p}\tilde{B}^{-1}_{p} \tilde{\beta}_{p}
          - \tilde{\beta}'_{p}\tilde{B}^{-1}_{p} \tilde{\beta}_{p}
          \bigg\}  \bigg]\\
          & \propto
           \exp\bigg[ -\frac{1}{2}
           \bigg\{ (\beta_{p} - \tilde{\beta}_{p})'
           \tilde{B}^{-1}_{p}(\beta_{p} - \tilde{\beta}_{p})
           \bigg\}  \bigg],
\end{align*}
where the last line follows by recognizing that
$\tilde{\beta}'_{p}\tilde{B}^{-1}_{p} \tilde{\beta}_{p}$ does not involve
$\beta_{p}$ and can therefore be absorbed in the constant of proportionality.
The result is the kernel of a Gaussian or normal density and hence
$\beta_{p}|z,w\sim N(\tilde{\beta}_{p}, \tilde{B}_{p})$.


Similar to the above approach, the full conditional distribution of $w$,
denoted by $\pi(w|z,\beta_{p})$ is proportional to $f(z|\beta_{p},w) \pi(w)$.
The kernel for each $w_{i}$ can be derived as,
\begin{align*}
\pi(w_{i}|z,\beta_{p})
          & \propto
          w_{i}^{-1/2} \exp\bigg[ -\frac{1}{2} \bigg(\frac{z_{i} - x'_{i}\beta_{p} - \theta w_{i}}
                  {\tau \sqrt{w_{i}}}   \bigg)^{2}  - w_{i}     \bigg]\\
          & \propto
          w_{i}^{-1/2} \exp\bigg[ -\frac{1}{2} \bigg( \frac{ (z_{i} - x'_{i}\beta_{p})^{2} + \theta^{2} w_{i}^{2}
          - 2\theta w_{i} (z_{i} - x'_{i}\beta_{p}) }{\tau^{2} w_{i}} + 2 w_{i}  \bigg)    \bigg]\\
          & \propto
          w_{i}^{-1/2} \exp\bigg[ -\frac{1}{2}  \bigg\{
          \frac{ (z_{i} - x'_{i}\beta_{p})^{2}}{\tau^{2}} \; w_{i}^{-1}
          + \bigg( \frac{\theta^{2}}{\tau^{2}} + 2 \bigg)
          w_{i} \bigg\}    \bigg]\\
          & \propto
          w_{i}^{-1/2} \exp\bigg[ -\frac{1}{2} \big\{ \tilde{\lambda_{i}} \, w_{i}^{-1} +
          \tilde{\eta} \, w_{i} \big\}
          \bigg].
\end{align*}
The last expression can be recognized as the kernel of the GIG distribution,
where,
\begin{equation*}
\tilde{\lambda_{i}} = \frac{ ( z_{i} -
x'_{i}\beta_{p} )^2 }{\tau^{2}}
                  \hspace{0.25in} \mathrm{and} \hspace{0.25in}
         \tilde{\eta} = \Big( \frac{\theta^2}{\tau^{2}} + 2 \Big).
\end{equation*}
Hence, as required $w_{i}|z,\beta_{p} \sim GIG(0.5, \tilde{\lambda_{i}},
\tilde{\eta})$.

The transformed cut-points $\delta_{p}$ do not have a tractable full
conditional density and hence sampled marginally of $(z,w)$ based on the full
likelihood \eqref{eq:likelihood}. The proposed values are generated from a
random-walk chain,
\begin{equation*}
\delta'_{p} = \delta_{p} + u,
\end{equation*}
where $u \sim N(0_{2}, \iota^{2} \hat{D})$, $\iota$ is a tuning parameter and
$\hat{D}$ denotes negative inverse Hessian, obtained by maximizing the
log-likelihood with respect to $\delta_{p}$. Given the the current value
$\delta_{p}$ and proposed value $\delta'_{p}$, the new value $\delta'_{p}$ is
accepted with MH probability,
\begin{equation*}
\alpha_{MH}(\delta_{p}, \delta'_{p}) = \min \bigg\{1,
\frac{ \;f(y|\beta_{p},\delta'_{p}) \;\pi(\beta_{p}, \delta'_{p})}
{f(y|\beta_{p},\delta_{p}) \;\pi(\beta_{p}, \delta_{p})}
\bigg\},
\end{equation*}
otherwise, the current value $\delta_{p}$ is repeated. The variance of $u$
may be tuned as needed for an appropriate step size and acceptance rate.


Finally, the full conditional density of the latent variable $z$ is a
truncated normal distribution where the cut-point vector $\gamma_{p}$ is
obtained based on one-to-one mapping from the transformed cut-point vector
$\delta_{p}$. Hence, $z$ is sampled as $z_{i}|y, \beta_{p},\gamma_{p},w$
$\sim$ $TN_{(\gamma_{p,j-1}, \gamma_{p,j})}(x'_{i}\beta_{p} + \theta w_{i},
\tau^{2}w_{i})$ for $i = 1, \cdots, n$.

\section{Conditional Densities in $\mathrm{OR_{II}}$ Model}\label{sec:appendixORII}

In the context of $\mathrm{OR_{II}}$ model, the complete posterior density
\eqref{eq:CompPostORII} is utilized to derive the full conditional densities
for the parameters of interest ($\beta_{p}$, $\sigma_{p}$, $\nu$) and the
latent variable $z$. The derivations follow the ordering as presented in
Algorithm~\ref{alg:algorithm2}.

The full conditional density of $\beta_{p}$ given by
$\pi(\beta_{p}|z,\sigma_{p},\nu)$ is proportional to $ \pi(\beta_{p}) \times
\allowbreak{f(z|\beta_{p},\sigma_{p}, \nu)}$ and its kernel can be written
as,
\begin{align*}
\pi(\beta_{p}|z,\sigma_{p},\nu)
          & \propto
          \exp\bigg[-\frac{1}{2} \bigg\{ \sum_{i=1}^{n}
          \bigg( \frac{z_{i} - x'_{i}\beta_{p} - \theta \nu_{i}}
          {\tau \sqrt{\sigma_{p} \nu_{i}}} \bigg)^2
          + (\beta_{p} - \beta_{p0} )' B_{p0}^{-1} (\beta_{p} - \beta_{p0} )
          \bigg\}  \bigg]\\
          & \propto
          \exp\bigg[-\frac{1}{2} \bigg\{ \beta_{p}'
          \bigg(\sum_{i=1}^{n} \frac{x_{i} x'_{i}}{\tau^{2} \sigma_{p} \nu_{i}}
          + B_{p0}^{-1}\bigg) \beta_{p}
          - \beta'_{p} \bigg( \sum_{i=1}^{n}
          \frac{x_{i}(z_{i}-\theta \nu_{i})}{\tau^{2} \sigma_{p} \nu_{i}}\\
          & \hspace{36pt} + B_{p0}^{-1} \beta_{p0}  \bigg)
          - \bigg( \sum_{i=1}^{n}
          \frac{x'_{i}(z_{i}-\theta \nu_{i})}{\tau^{2} \sigma_{p} \nu_{i}}
           + \beta'_{p0} B_{p0}^{-1}  \bigg) \beta_{p}
           \bigg\}  \bigg]\\
          & \propto
          \exp\bigg[ -\frac{1}{2} \bigg\{ \beta'_{p}\tilde{B}^{-1}_{p}\beta_{p}
          - \beta'_{p}\tilde{B}^{-1}_{p}\tilde{\beta}_{p}
          - \tilde{\beta}'_{p} \tilde{B}^{-1}_{p}\beta_{p}
              \bigg\}  \bigg],
\end{align*}
where as done earlier, the second line omits all terms not involving
$\beta_{p}$ and the third line uses the terms $\tilde{B}_{p}$ and
$\tilde{\beta}_{p}$, which are defined as follows,
\begin{equation*}
\tilde{B}^{-1}_{p} = \bigg(\sum_{i=1}^{n} \frac{x_{i} x'_{i}}{\tau^{2}
         \sigma_{p} \nu_{i}} + B_{p0}^{-1} \bigg)
         \hspace{0.25in} \mathrm{and} \hspace{0.25in}
         \tilde{\beta}_{p} =
         \tilde{B}_{p}\bigg( \sum_{i=1}^{n} \frac{x_{i}(z_{i}
         - \theta \nu_{i})}{\tau^{2} \sigma_{p} \nu_{i}}
         + B_{p0}^{-1} \beta_{p0} \bigg).
\end{equation*}
Note that ($\tilde{B}_{p}$, $\tilde{\beta}_{p}$) are different compared to
that of \ref{sec:appendixORI}. Adding and subtracting $\tilde{\beta}'_{p}
\tilde{B}^{-1}_{p}\tilde{\beta}_{p}$ inside the curly braces, the square can
be completed as,
\begin{align*}
\pi(\beta_{p}|z,\sigma_{p},\nu) & \propto
          \exp\bigg[ -\frac{1}{2} \bigg\{ \beta'_{p}\tilde{B}^{-1}_{p}\beta_{p}
          - \beta'_{p}\tilde{B}^{-1}_{p}\tilde{\beta}_{p}
          - \tilde{\beta}'_{p}\tilde{B}^{-1}_{p}\beta_{p}
          + \tilde{\beta}'_{p}\tilde{B}^{-1}_{p} \tilde{\beta}_{p}
          - \tilde{\beta}'_{p}\tilde{B}^{-1}_{p} \tilde{\beta}_{p}
          \bigg\}  \bigg]\\
          & \propto
           \exp\bigg[ -\frac{1}{2}
           \bigg\{ (\beta_{p} - \tilde{\beta}_{p})' \tilde{B}_{p}^{-1}
           (\beta_{p} - \tilde{\beta}_{p})
           \bigg\}  \bigg],
\end{align*}
where again the last line follows by recognizing that
$\tilde{\beta}'_{p}\tilde{B}^{-1}_{p} \tilde{\beta}_{p}$ does not involve
$\beta_{p}$ and can therefore be absorbed in the constant of proportionality.
The result is the kernel of the Gaussian or normal density and hence
$\beta_{p}|z,\sigma_{p},\nu \sim N(\tilde{\beta}_{p}, \tilde{B}_{p})$.

The full conditional density of scale parameter $\sigma_{p}$, represented by
$\pi(\sigma_{p}|z,\beta_{p},\nu)$ is proportional to
$f(z|\beta_{p},\nu,\sigma_{p}) \,\pi(\nu|\sigma_{p})\,\pi(\sigma_{p})$, and
can be derived as follows ,
\begin{align*}
\pi(\sigma_{p}|z,\beta_{p},\nu)
          & \propto
          \prod_{i=1}^{n} \bigg\{ \sigma_{p}^{-1/2} \exp \bigg[-\frac{1}{2}
          \bigg( \frac{z_{i} - x'_{i}\beta_{p} - \theta \nu_{i}}
          {\tau \sqrt{\sigma_{p} \nu_{i}}} \bigg)^2   \bigg]
          \times \sigma_{p}^{-1} \exp\bigg(-\frac{\nu_{i}}{\sigma_{p}} \bigg)
          \bigg\}\\
          &  \hspace{30pt} \times \exp\bigg[-\frac{d_{0}}{2\sigma_{p}} \bigg]  \sigma_{p}^{-(n_{0}/2 + 1)}\\
          & \propto
          \sigma_{p}^{-\big(\frac{n_{0}}{2}+\frac{3n}{2} + 1 \big)} \exp\bigg[ -\frac{1}{\sigma_{p}}
          \bigg\{ \sum_{i=1}^{n} \frac{(z_{i} - x'_{i}\beta_{p} - \theta \nu_{i})^{2}}{2 \tau^{2} \nu_{i}}
          + \frac{d_{0}}{2}  + \sum_{i=1}^{n} \nu_{i}    \bigg\}     \bigg].
\end{align*}
The last expression can be recognized as the kernel of an inverse-gamma
distribution, where,
\begin{equation*}
\tilde{n} = (n_{0} + 3n)  \hspace{0.25in} \mathrm{and} \hspace{0.25in}
         \tilde{d} = \sum_{i=1}^{n}(z_{i} - x'_{i}\beta_{p} - \theta \nu_{i})^{2}/\tau^{2}\nu_{i}
         + d_{0} + 2 \sum_{i=1}^{n} \nu_{i}.
\end{equation*}
Therefore, $\sigma_{p}|z,\beta_{p},\nu \sim IG(\tilde{n}/2, \tilde{d}/2)$.

The full conditional density of $\nu$, unlike $\beta_{p}$ and $\sigma_{p}$,
is not a simple update of its prior distribution. The full conditional
distribution $\pi(\nu|z,\beta_{p},\sigma_{p})$ is proportional to
$f(z|\beta_{p},\nu,\sigma_{p}) \,\pi(\nu)$ and the kernel for each $\nu_{i}$
can be derived as,
\begin{align*}
\pi(\nu_{i}|z,\beta_{p},\sigma_{p})
          & \propto
          \nu_{i}^{-1/2} \exp\bigg[ -\frac{1}{2} \bigg(\frac{z_{i} - x'_{i}\beta_{p} - \theta \nu_{i}}
                  {\tau \sqrt{\sigma_{p} \nu_{i}}}   \bigg)^{2}  - \frac{\nu_{i}}{\sigma_{p}}     \bigg]\\
          & \propto
          \nu_{i}^{-1/2} \exp\bigg[ -\frac{1}{2\sigma_{p}} \bigg( \frac{ (z_{i} - x'_{i}\beta_{p})^{2} +
          \theta^{2} \nu_{i}^{2}
          - 2\theta \nu_{i} (z_{i} - x'_{i}\beta_{p}) }{\tau^{2} \nu_{i}} + 2\nu_{i}  \bigg)    \bigg]\\
          & \propto
          \nu_{i}^{-1/2} \exp\bigg[ -\frac{1}{2}  \bigg\{
          \frac{ (z_{i} - x'_{i}\beta_{p})^{2}}{\tau^{2}\sigma_{p}} \; \nu_{i}^{-1} +
           \bigg( \frac{\theta^{2}}{\tau^{2} \sigma_{p}} + \frac{2}{\sigma_{p}} \bigg) \nu_{i}
           \bigg\}    \bigg]\\
          & \propto
          \nu_{i}^{-1/2} \exp\bigg[ -\frac{1}{2} \big\{ \tilde{\lambda_{i}} \, \nu_{i}^{-1} +
          \tilde{\eta} \, \nu_{i} \big\}   \bigg].
\end{align*}
The last expression can be recognized as the kernel of the GIG distribution,
where,
\begin{equation*}
\tilde{\lambda_{i}} = \frac{ ( z_{i} - x'_{i}\beta_{p} )^2 }{\tau^{2}\sigma_{p}}
                  \hspace{0.25in} \mathrm{and} \hspace{0.25in}
         \tilde{\eta} = \Big( \frac{\theta^2}{\tau^{2} \sigma_{p}} + \frac{2}{\sigma_{p}} \Big).
\end{equation*}
Hence, as required $\nu_{i}|z,\beta_{p},\sigma_{p} \sim GIG(0.5,
\tilde{\lambda_{i}}, \tilde{\eta})$. Note that the definitions of
$\tilde{\lambda_{i}}$ and $\tilde{\eta}$ are different compared to that of
\ref{sec:appendixORI}.


Lastly, the full conditional density of the latent variable $z$ is a
truncated normal distribution and sampled as $z_{i}|y,
\beta_{p},\gamma,\sigma_{p}, \nu$ $\sim$ $TN_{(\gamma_{j-1},
\gamma_{j})}(x'_{i}\beta_{p} + \theta \nu_{i}, \tau^{2}\sigma_{p} \nu_{i})$
for $i = 1, \cdots, n$ and $j=1,2,3$. Note that for the $\mathrm{OR_{II}}$
model, the cut-point vector $\gamma$ is completely known.


\bibliographystyle{ba}
\bibliography{bqrorBibliography}

\begin{acknowledgement}
The paper is dedicated to Ivan Jeliazkov and Dale Poirier for introducing me
to Bayesian econometrics. Special thanks to the editor and two anonymous
referees for helpful comments and suggestions. The paper forms a chapter of
my dissertation and was largely written at the University of California,
Irvine, USA.
\end{acknowledgement}

\end{document}